\documentclass[11pt,a4paper]{article}

\usepackage{amsmath,amssymb,cite,epsfig}

\begin{document}

\title{Thin--shell wormholes with a double layer in quadratic $F(R)$ gravity} 
\author{Ernesto F. Eiroa$^{1, 2}$\thanks{e-mail: eiroa@iafe.uba.ar}, Griselda Figueroa Aguirre$^{1}$\thanks{e-mail: gfigueroa@iafe.uba.ar}\\
{\small $^1$ Instituto de Astronom\'{\i}a y F\'{\i}sica del Espacio (IAFE, CONICET-UBA),}\\
{\small Casilla de Correo 67, Sucursal 28, 1428, Buenos Aires, Argentina}\\
{\small $^2$ Departamento de F\'{\i}sica, Facultad de Ciencias Exactas y 
Naturales,} \\ 
{\small Universidad de Buenos Aires, Ciudad Universitaria Pabell\'on I, 1428, 
Buenos Aires, Argentina}} 

\maketitle

\begin{abstract}

We present a family of spherically symmetric Lorentzian wormholes in quadratic $F(R)$ gravity, with a thin shell of matter corresponding to the throat. At each side of the shell the geometry has a different constant value of the curvature scalar $R$. The junction conditions determine the equation of state between the pressure and energy density at the throat, where a double layer is also located. We analyze the stability of the configurations under perturbations preserving the spherical symmetry. In particular, we study thin--shell wormholes with mass and charge. We find that there exist values of the parameters for which stable static solutions are possible.\\

\noindent 
PACS number(s): 04.20.Gz, 04.40.Nr, 98.80.Jk

\end{abstract}

\section{Introduction}\label{intro} 

Traversable wormholes \cite{motho,visser} are spacetimes characterized by the presence of a throat connecting two regions of the same universe or two different universes. At the throat the geometry opens up, which is reflected in the so called flare--out condition. In General Relativity, these theoretical objects require exotic matter (i.e. a fluid that does not satisfies the null energy condition) at least in the region close to the throat \cite{motho,visser,hovis}. The exotic matter can be reduced to an arbitrary small amount \cite{viskardad}, but at the cost of large pressures at the throat \cite{lst}. The junction conditions in General Relativity \cite{daris} are used to join two solutions across a hypersurface in different contexts, for example the interior and exterior solutions corresponding to stars or galaxies, in the study of thin layers of matter, in cosmological models, etc. Wormholes can be also constructed \cite{visser} by a cut and paste procedure, with a thin--shell that corresponds to the throat in most cases. This class of wormholes has received great attention in the literature due to its simplicity and because the exotic matter can be confined to the shell. The junction formalism can also be adopted to have a suitable asymptotic behavior or to reduce the need of exotic matter for wormholes with a continuous stress--energy tensor at the throat. In highly symmetric configurations the stability analysis of thin shells can be performed quite easily. Many studies of spherically symmetric thin--shell wormholes, with a linearized equation of state at the throat, have been done by taking radial perturbations (e.g. \cite{poisson,isla-eirom,eir-lmv,dilem}). Non--linear equations of state at the shell were used to model the exotic fluid threading the throat by several authors \cite{chaplywh1,chaplywh2,chaplywh3,varela}. Cylindrically symmetric thin--shell wormholes were considered in recent years (e.g. \cite{cil}). Thin--shell wormholes were also analyzed within Brans-Dicke gravity \cite{whbd}, a well known alternative theory to General Relativity. 

The standard model of cosmology requires the presence of dark matter and dark energy, in order to understand the observed features of the Universe, i.e. the galaxy rotation curves, the anisotropy of the microwave background radiation, and the current accelerated expansion. Another important issue is the inflationary epoch, which is usually explained by adopting a scalar field. Modifications of General Relativity were introduced in the literature with the purpose of avoiding the presence of these non standard fluids. One of them is $F(R)$ gravity \cite{revfr}, in which a function $F(R)$ of the Ricci scalar $R$ replaces the Einstein--Hilbert Lagrangian in the gravitational action. This theory can provide a unified picture of both inflation in the early Universe and the accelerated expansion at later times. It is also of interest the study of compact objects within $F(R)$ gravity; in particular, static and spherically symmetric black hole solutions \cite{bhfr1,bhfr2,bhfr3} and traversable wormholes \cite{whfr1,whfr2} were investigated in recent years. The junction conditions have been extended to $F(R)$ gravity in the last decade \cite{dss,js1}; these conditions are more demanding in non--linear $F(R)$ theories than in General Relativity, because they always require continuity of the trace of the second fundamental form at the shell and, except in the case of quadratic $F(R)$, the continuity of the curvature scalar $R$. In quadratic $F(R)$, the Ricci scalar can be discontinuous at the matching hypersurface and, as a consequence, the shell will have, besides the standard energy--momentum tensor, an external energy flux vector, an external scalar tension (or pressure), and another energy--momentum contribution resembling classical dipole distributions \cite{js2,js3}, which can be interpreted as a gravitational double layer. All these contributions are necessary to make the whole energy--momentum tensor divergence free \cite{js2,js3}. The junction conditions obtained in quadratic $F(R)$ were extended in the last year to the most general gravitational theory with a Lagrangian quadratic in the curvature \cite{js4}.

Thin--shell wormholes in $F(R)$ gravity, symmetric across the throat and with a constant curvature scalar, were recently studied in \cite{gfa}. Here we consider thin--shell wormholes with spherical symmetry in quadratic $F(R)$ theory, with constant and different Ricci scalars at each side of the throat; we analyze the characteristics of the fluid at the shell and we investigate their stability under radial perturbations. In Sec. \ref{tswh}, we perform the wormhole construction for a large class of possible geometries. In Sec. \ref{stab}, we study the stability of the static configurations. In Sec. \ref{charge}, we apply the formalism to charged wormholes. Finally, in Sec. \ref{conclu}, we present the conclusions of the paper. We take units so that $G=c=1$, with $G$ the gravitational constant and $c$ the speed of light.

\section{Wormhole construction}\label{tswh}

In this work, we start from two spherically symmetric geometries of the form 
\begin{equation} 
ds_{1,2}^2=-A_{1,2} (r) dt^2+A_{1,2} (r)^{-1} dr^2+r^2(d\theta^2 + \sin^2\theta d\varphi^2),
\label{metric}
\end{equation}
where $r>0$ is the radial coordinate, $0\le \theta \le \pi$, and $0\le \varphi<2\pi $ are the angular coordinates. This class of static spacetimes in $F(R)$ theories includes (in the metric formalism) several well known exact solutions, among them those with constant curvature scalar $R$, corresponding to vacuum \cite{bhfr1,bhfr2} and to a Maxwell field in 3+1 dimensions \cite{bhfr2} (for other metrics of this form, see also \cite{bhfr3}). For the construction of wormholes, we use the thin--shell formalism in $F(R)$ gravity. We choose a radius $a$, larger than the event horizons of both metrics (in case they have them), we cut from each geometry the region with $r\geq a$:
\begin{equation} 
\mathcal{M}^{1,2 }=\{X_{1,2}^{\alpha }=(t,r,\theta,\varphi)/r\geq a\},  
\end{equation}
and paste them at the hypersurface
\begin{equation} 
\Sigma \equiv \Sigma ^{1,2 }=\{X_{1,2}/G(r)=r-a=0\}, 
\end{equation}
in order to obtain a new geodesically complete manifold $\mathcal{M}=\mathcal{M}^{1} \cup \mathcal{M}^{2}$. The indexes $1$ and $2$ used along the present work correspond, respectively, to $\mathcal{M}^{1}$ and $\mathcal{M}^{2}$. The flare-out condition is satisfied because the area  $4\pi r^2$ is minimal when $r=a$; therefore, the manifold $\mathcal{M}$ represents a wormhole with two regions connected by a throat of radius $a$. We can define a global radial coordinate $\mathcal{M}$ by using the proper radial distance: $\ell =\pm \int_{a}^{r}\sqrt{1/A_{1,2} (r)}dr$, where the ($+$) sign corresponds to  $\mathcal{M}^{1}$ and the ($-$) sign to $\mathcal{M}^{2}$, and the throat is placed in $\ell = 0$. We denote the jump across the shell of any quantity $\Upsilon  $ by $[\Upsilon ]\equiv (\Upsilon ^{1}-\Upsilon  ^{2})|_\Sigma $, the unit normals to $\Sigma $ in $\mathcal{M}$ by $n_{\gamma }^{1,2 }$, the first fundamental form by $h_{\mu \nu}$, and the second fundamental form  (extrinsic curvature) by $K_{\mu \nu}$.

In  $F(R)$ theories there exists an additional condition \cite{js1} besides the continuity of the first fundamental form across the shell  ($[h_{\mu \nu }]=0$), which is the continuity of the trace of the second fundamental form ($[K^{\mu}_{\;\; \mu}]=0$). If $F'''(R) \neq 0$ (the prime represents the derivative with respect to $R$), the continuity of $R$ across the shell ($[R]=0$) is also required as a third condition  \cite{js1}. However, in the quadratic theory $F(R)=R -2\Lambda +\alpha R^2$, for which $F'''(R) = 0$, it is possible to have the discontinuity of $R$ at $\Sigma $. In our construction, we take two metrics with different values of $R$ at each side of the throat, therefore $[R]\neq0$. The field equations at the shell in the quadratic case read \cite{js1,js4}
\begin{equation}
\kappa S_{\mu \nu} =-[K_{\mu\nu}]+2\alpha( [n^{\gamma }\nabla_{\gamma}R] h_{\mu\nu}-[RK_{\mu\nu}]), \;\;\;\; n^{\mu}S_{\mu\nu}=0,
\label{LanczosQuad}
\end{equation}
where $\kappa =8\pi $ and $S_{\mu \nu}$ represents the energy-momentum tensor at the shell; along with
\begin{equation}
\kappa\mathcal{T}_\mu=-2\alpha\nabla_\mu[R], \;\;\;\; n^{\mu}\mathcal{T}_\mu=0,
\label{Tmu}
\end{equation}
\begin{equation}
\kappa\mathcal{T}=2\alpha [R] K^\gamma{}_\gamma ,
\label{Tg}
\end{equation}
and a two-covariant symmetric tensor distribution 
\begin{equation}
\kappa \mathcal{T}_{\mu \nu}=\nabla_{\gamma } \left( 2\alpha [R] h_{\mu \nu } n^{\gamma } \delta ^{\Sigma }\right),
\label{dualay1}
\end{equation}
where $\delta ^{\Sigma }$ is the Dirac delta with support on $\Sigma $, or equivalently 
\begin{equation}
\kappa \left<\mathcal{T}_{\mu \nu},\Psi ^{\mu \nu } \right> = -\int_\Sigma 2\alpha[R] h_{\mu \nu }  n^\gamma\nabla_\gamma \Psi ^{\mu \nu },
\label{dualay2}
\end{equation}
for any test tensor field $\Psi ^{\mu \nu }$. In quadratic $F(R)$, besides the standard energy-momentum tensor $S_{\mu \nu}$, the shell has an external energy flux vector $\mathcal{T} _{\mu}$, an external scalar pressure/tension $\mathcal{T} $, and a double layer energy-momentum distribution $\mathcal{T}_{\mu \nu }$. This Dirac ``delta prime'' type contribution has a resemblance with classical dipole distributions  \cite{js1,js4}. The ``dipole'' distribution $\mathcal{T}_{\mu \nu }$ has a strength $\kappa \mathcal{P}_{\mu \nu } =2\alpha[R] h_{\mu \nu }$, which satisfies $ \mathcal{P}_{\mu \nu } = \mathcal{P}_{\nu \mu }$ and $n^{\mu } \mathcal{P}_{\nu \mu } =0$. All these contributions are necessary to make the complete energy-momentum tensor divergence free, so that it is locally conserved \cite{js1,js4}.

We adopt constant curvature scalars, but with different values for each region of $\mathcal{M}$, i.e. $R_1 \neq R_2$, so from Eq. (\ref{LanczosQuad}) we obtain
\begin{equation}
\kappa S_{\mu \nu} =-[K_{\mu\nu}]-2\alpha[RK_{\mu\nu}].
\label{LanczosGen2}
\end{equation}
At the surface $\Sigma $, we define the coordinates $\xi ^{i}=(\tau ,\theta,\varphi )$, with $\tau $ the proper time on the shell. The throat radius is a function of the proper time: $a(\tau)$.  The first fundamental form associated with the two sides of the shell is given by
\begin{equation}
h^{1,2}_{ij}= \left. g^{1,2}_{\mu\nu}\frac{\partial X^{\mu}_{1,2}}{\partial\xi^{i}}\frac{\partial X^{\nu}_{1,2}}{\partial\xi^{j}}\right| _{\Sigma },
\end{equation}
and the second fundamental form is calculated from
\begin{equation}
K_{ij}^{1,2 }=-n_{\gamma }^{1,2 }\left. \left( \frac{\partial ^{2}X^{\gamma
}_{1,2} } {\partial \xi ^{i}\partial \xi ^{j}}+\Gamma _{\alpha \beta }^{\gamma }
\frac{ \partial X^{\alpha }_{1,2}}{\partial \xi ^{i}}\frac{\partial X^{\beta }_{1,2}}{
\partial \xi ^{j}}\right) \right| _{\Sigma },
\label{sff}
\end{equation}
where
\begin{equation}
n_{\gamma }^{1,2 }=\pm \left\{ \left. \left| g^{\alpha \beta }_{1,2}\frac{\partial G}{\partial
X^{\alpha }_{1,2}}\frac{\partial G}{\partial X^{\beta }_{1,2}}\right| ^{-1/2}
\frac{\partial G}{\partial X^{\gamma }_{1,2}} \right\} \right| _{\Sigma }.
\end{equation}
For the metrics given by Eq. (\ref{metric}), the unit normals ($n^{\gamma }n_{\gamma }=1$) read
\begin{equation}
n_{\gamma }^{1,2 }=\pm \left(-\dot{a},\frac{\sqrt{A_{1,2}(a)+\dot{a}^2}}{A_{1,2}(a)},0,0 \right),
\end{equation}
with the dot representing the derivative with respect to $\tau$. We adopt at the shell the orthonormal basis $\{ e_{\hat{\tau}}=e_{\tau }, e_{\hat{\theta}}=a^{-1}e_{\theta }, e_{\hat{\varphi}}=(a\sin \theta )^{-1} e_{\varphi }\} $, then we have that the first fundamental form is $h^{1,2}_{\hat{\imath}\hat{\jmath}}= \mathrm{diag}(-1,1,1)$, and the second fundamental form has non null components
\begin{equation} 
K_{\hat{\theta}\hat{\theta}}^{1,2 }=K_{\hat{\varphi}\hat{\varphi}}^{1,2
}=\pm \frac{1}{a}\sqrt{A_{1,2} (a) +\dot{a}^2},
\label{Kth}
\end{equation}
and
\begin{equation} 
K_{\hat{\tau}\hat{\tau}}^{1,2 }=\mp\left(  \frac{\ddot{a}}{\sqrt{A_{1,2} (a) +\dot{a}^2}}+\frac{1}{2}\frac{A'_{1,2}(a)}{\sqrt{A_{1,2}(a) +\dot{a}^2}} \right),
\label{Ktau}
\end{equation}
where the prime represents the derivative with respect to $r$. By using Eqs. (\ref{Kth}) and (\ref{Ktau}), the condition $[K^{\hat{\imath}}_{\;\; \hat{\imath}}]=0$ adopts the form
\begin{equation} 
\frac{2a\ddot{a}+a A_{1}'(a)+4 \left(A_{1}(a) +\dot{a}^2\right)}{\sqrt{A_{1}(a) +\dot{a}^2}}+\frac{2a\ddot{a}+a A_{2}'(a)+4 \left(A_{2}(a) +\dot{a}^2\right)}{\sqrt{A_{2}(a) +\dot{a}^2}}=0.
\label{CondGen}
\end{equation}
In Eq. (\ref{LanczosGen2}) we replace the surface stress-energy tensor $S_{\hat{\imath}\hat{\jmath} }={\rm diag}(\sigma ,p_{\hat{\theta}},p_{\hat{\varphi}})$, with $\sigma$ the surface energy density and $p_{\hat{\theta}}$, $p_{\hat{\varphi}}$ the transverse pressures, and we obtain 
\begin{equation} 
\sigma= \frac{2 \ddot{a}+A'_{1}(a)}{2\kappa\sqrt{A_{1}(a)+\dot{a}^2}}\left(1+2\alpha R_{1}\right)+\frac{2 \ddot{a}+A'_{2}(a)}{2\kappa\sqrt{A_{2}(a)+\dot{a}^2}}\left(1+2\alpha R_{2}\right) ,
\label{enden}
\end{equation}
and
\begin{equation}
p=-\frac{\sqrt{A_{1}(a)+\dot{a}^2}}{\kappa a} \left(1+2\alpha R_{1} \right)-\frac{\sqrt{A_{2}(a)+\dot{a}^2}}{\kappa a} \left(1+2\alpha R_{2} \right),
\label{pres}
\end{equation}
where $p=p_{\hat{\theta}}=p_{\hat{\varphi}}$. From Eq. (\ref{Tmu}) we see that $\mathcal{T}_\mu =0$. The external scalar tension/pressure $\mathcal{T}$ is given by 
\begin{equation}
\mathcal{T}=\frac{2\alpha [R]}{\kappa \sqrt{A_{1}(a)+\dot{a}^2}}\left(\ddot{a}+\frac{A'_{1}(a)}{2}+\frac{2}{a}\left(A_{1}(a)+\dot{a}^2\right)\right),
\label{T}
\end{equation}
which by using Eq. (\ref{CondGen}) can be rewritten in the form
\begin{equation} 
\mathcal{T}=\frac{ 2a\ddot{a}+a A_{1}'+4 \left(A_{1}(a) +\dot{a}^2\right)}{\kappa a\sqrt{A_{1}(a) +\dot{a}^2}}\alpha R_{1}+\frac{2a\ddot{a}+a A_{2}'+4 \left(A_{2}(a) +\dot{a}^2\right)}{\kappa a\sqrt{A_{2}(a) +\dot{a}^2}}\alpha R_{2}.
\label{Trewrit}
\end{equation}
It is easy to see from Eqs. (\ref{enden}), (\ref{pres}), and (\ref{Trewrit}) that there is an equation of state that relates $\sigma$, $p$, and $\mathcal{T}$
\begin{equation}
\sigma-2p=\mathcal{T}.
\label{state}
\end{equation}
By taking the time derivative of Eq. (\ref{state}) and using that $\dot{p}= - (\dot{a}/a) (\sigma + p)$, we find the generalized continuity equation
\begin{equation} 
\dot{\sigma}+\frac{2\dot{a}}{a}(\sigma + p) = \dot{\mathcal{T}},
\label{continuity}
\end{equation}
or equivalently 
\begin{equation} 
\frac{d}{d\tau}(\mathcal{A}\sigma )+ p \frac{d\mathcal{A}}{d\tau} = \mathcal{A} \frac{d\mathcal{T}}{d\tau},
\label{continuity-bis}
\end{equation}
where $\mathcal{A}=4\pi a^2$ is the area of the shell. In the left hand side of this equation, the first term can be interpreted as the change in the total energy of the throat, the second one as the work done by the internal pressure, while the right hand side represents an external flux. The dual layer distribution $\mathcal{T}_{\mu \nu }$, obtained from Eq. (\ref{dualay2}), should satisfy
\begin{equation} 
\langle \mathcal{T}_{\mu \nu } ,\Psi ^{\mu \nu } \rangle= - \int_{\Sigma}  \mathcal{P}_{\mu \nu } \left( n^t \nabla _t \Psi ^{\mu \nu} + n^r \nabla _r \Psi ^{\mu \nu } \right),
\label{dualay3}
\end{equation}
for any test tensor field $\Psi ^{\mu \nu }$. The dual layer distribution strength, in the orthonormal basis, has components 
\begin{equation}
- \mathcal{P}_{\tau \tau} =\mathcal{P}_{\hat{\theta}\hat{\theta} }=\mathcal{P}_{\hat{\varphi}\hat{\varphi} }=2\alpha[R]/\kappa .
\label{dlsob}
\end{equation}
Note that these components only depend on $\alpha $ and $[R]$, so the dependence of $\mathcal{T}_{\hat{\imath}\hat{\jmath} }$ on the particular form of the metric is through the unit normal and the covariant derivative.

\section{Static configurations: stability}\label{stab}

The condition resulting from Eq. (\ref{CondGen}) for static wormholes is reduced to the form
\begin{equation} 
\frac{ a_0 A'_{1}(a_0)+4A_{1}(a_0)}{\sqrt{A_{1}(a_0)}}+\frac{ a_0 A'_{2}(a_0)+4A_{2}(a_0)}{\sqrt{A_{2}(a_0)}}=0,
\label{CondEstatico}
\end{equation}
where $a_0$ is the radius corresponding to the throat. The surface energy density, the pressure, and the external tension/pressure in the static case take the form, respectively,
\begin{equation} 
\sigma_0= \frac{A'_{1}(a_{0})}{2\kappa\sqrt{A_{1}(a_{0})}}\left(1+2\alpha R_{1}\right)+\frac{A'_{2}(a_{0})}{2\kappa\sqrt{A_{2}(a_{0})}}\left(1+2\alpha R_{2}\right) ,
\label{enden0}
\end{equation}
\begin{equation}
p_0=-\frac{\sqrt{A_{1}(a_{0})}}{\kappa a_{0}} \left(1+2\alpha R_{1} \right)-\frac{\sqrt{A_{2}(a_{0})}}{\kappa a_{0}} \left(1+2\alpha R_{2} \right),
\label{pres0}
\end{equation}
and
\begin{equation}
\mathcal{T}_0=\frac{ a_0 A_{1}'(a_0) + 4 A_{1}(a_0)}{\kappa a_0 \sqrt{A_{1}(a_0)}}\alpha R_{1} +\frac{a_0 A_{2}'(a_0)+4 A_{2}(a_0)}{\kappa a_0 \sqrt{A_{2}(a_0)}}\alpha R_{2}.
\label{T0}
\end{equation}
The equation of state becomes $\sigma_0 -2 p_0 =  \mathcal{T}_0$. The static dual layer distribution resulting from Eq. (\ref{dualay3}) has to fulfill
\begin{equation} 
\langle \mathcal{T}_{\mu \nu } ,\Psi ^{\mu \nu } \rangle=  - \int_{\Sigma}  \mathcal{P}_{\mu \nu } n^r \nabla _r \Psi ^{\mu \nu },
\label{dualay4}
\end{equation}
for an arbitrary tensor field $\Psi ^{\mu \nu }$. The components in the orthonormal frame $\mathcal{P}_{\hat{\imath}\hat{\jmath} }$ of the distribution strength are given by Eq. (\ref{dlsob}).

For the stability analysis of the static configurations under radial perturbations, we extend the General Relativity procedure \cite{poisson} to quadratic $F(R)$ gravity. By using that $\ddot{a}= (1/2)d(\dot{a}^2)/da$ and defining $z=\sqrt{A_{1}(a) +\dot{a}^2}+\sqrt{A_{2}(a) +\dot{a}^2}$, we can rewrite Eq. (\ref{CondGen}) into 
\begin{equation}
a z'+2 z=0.
\label{CondGen_z}
\end{equation}
By integrating Eq. (\ref{CondGen_z}) we obtain
\begin{equation}
 \sqrt{A_{1}(a) +\dot{a}^2}+\sqrt{A_{2}(a) +\dot{a}^2} = \frac{a_0^2}{a^2} \left( \sqrt{A_1(a_0)} + \sqrt{A_2(a_0)} \right),
\label{soledif}
\end{equation}
from which it is possible to express the dynamics of the throat in the form
\begin{equation}
\dot{a}^{2}=-V(a),
\label{condicionPot}
\end{equation}
where 
\begin{equation}
V(a)= -\frac{a_{0}^4\left(\sqrt{A_{1}(a_{0})}+\sqrt{A_{2}(a_{0})}\right)^2}{4a^4} +\frac{A_{1}(a)+A_{2}(a)}{2} -\frac{a^4 \left(A_{1}(a)-A_{2}(a)\right)^{2}}{4 a_{0}^4\left(\sqrt{A_{1}(a_{0})}+\sqrt{A_{2}(a_{0})}\right)^2}
\label{potencial}
\end{equation}
can be understood as a potential. It is not difficult to verify that $V(a_0)=0$ and, with the aid of Eq. (\ref{CondEstatico}), that also $V'(a_0)=0$, while the second derivative of the potential at $a_0$ takes the form
\begin{eqnarray}
V''(a_0)&=& -\frac{5 \left(\sqrt{A_{1}(a_{0})}+\sqrt{A_{2}(a_{0}})\right)^{2}}{a_{0}^2}+\frac{A_{1}''(a_{0})+A_{2}''(a_{0})}{2}\nonumber \\
&& -\frac{3\left( A_{1}(a_{0})-A_{2}(a_{0})\right) ^2}{a_{0}^2\left(\sqrt{A_{1}(a_{0})}+\sqrt{A_{2}(a_{0}})\right)^{2}}
-\frac{4\left(A_{1}(a_{0})-A_{2}(a_{0})\right)\left(A_{1}'(a_{0})-A_{2}'(a_{0})\right)}{a_{0}\left(\sqrt{A_{1}(a_{0})}+\sqrt{A_{2}(a_{0})}\right)^{2}}  \nonumber \\
&& -\frac{\left(A_{1}'(a_{0})-A_{2}'(a_{0})\right)^2}{2\left(\sqrt{A_{1}(a_{0})}+\sqrt{A_{2}(a_{0})}\right)^{2}}-\frac{\left(A_{1}(a_{0})-A_{2}(a_{0})\right)\left(A_{1}''(a_{0})-A_{2}''(a_{0})\right)}{2\left(\sqrt{A_{1}(a_{0})}+\sqrt{A_{2}(a_{0})}\right)^{2}}.
\label{potencial2der}
\end{eqnarray}
The condition for a configuration with radius $a_0$ to be stable under radial perturbations is that $V''(a_0)>0$.

\section{Charged wormholes}\label{charge}

As an application of the formalism introduced above, we construct charged wormholes with different values of the curvature scalar at each side of the throat. The action for quadratic $F(R)$ in the presence of the electromagnetic field  tensor  $\mathcal{F}_{\mu \nu }=\partial _{\mu }\mathcal{A}_{\nu } -\partial _{\nu }\mathcal{A}_{\mu }$ has the form
\begin{equation}
S=\frac{1}{2 \kappa}\int d^4x \sqrt{|g|} (R -2\Lambda +\alpha R^2-\mathcal{F}_{\mu\nu}\mathcal{F}^{\mu\nu}),
\label{action} 
\end{equation} 
where $g=\det (g_{\mu \nu})$. In the metric formalism, the field equations obtained from the action above, for an electromagnetic potential $\mathcal{A}_{\mu}=(\mathcal{V}(r),0,0,0)$ and a constant curvature scalar $R$ admit the spherically symmetric solution given by Eq. (\ref{metric}), in which the metric function reads  \cite{bhfr2} 
\begin{equation} 
A(r) = 1-\frac{2M}{r}+\frac{Q^2}{ (1+2\alpha R) r^2}-\frac{R r^2}{12},
\label{A-metric}
\end{equation}
where $M$ is the mass and $Q$ is the charge. In this solution, $\mathcal{V}(r)=Q/r$ and the constant value of $R$ is related to the cosmological constant by $R=4\Lambda $. The metric has some relevant aspects to notice: the squared charge $Q^2$ is corrected by a factor $1/(1+2\alpha R)$ with respect to the General Relativity solution; the geometry is singular at $r=0$; the position of the horizons, determined by the zeros of $A(r)$, are given by the positive roots of a fourth degree polynomial. For $R>0$, there exist three horizons for small values of $|Q|$: the inner one with radius $r_i$, the event one with radius $r_h$, and the cosmological one with radius $r_c$; when the charge reaches a critical value $|Q|=Q_c$, the inner and the event horizons fuse into one; when $|Q|>Q_c$ there is a naked singularity at the origin and only the cosmological horizon remains. When $R<0$, for small values of $|Q|$ there are two horizons: the inner one with radius $r_i$ and the event one with radius $r_h$; when the charge is $|Q|=Q_c$, both horizons fuse into one, and if $|Q|>Q_c$ there is only a naked singularity. 

\begin{figure}[t!]
\centering
\includegraphics[width=\textwidth]{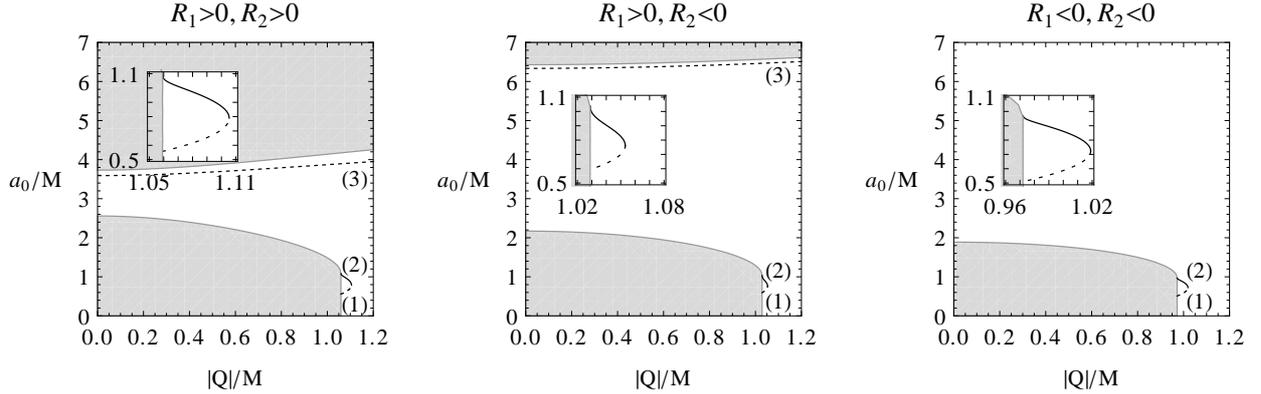}
\caption{Static solutions with radius $a_0$ corresponding to wormholes with $|R_1| M^2=0.2$ and $|R_2| M^2=0.4$, for different combinations of the signs of $R_1$ and $R_2$; $M$ and $Q$ are the mass and charge, respectively. The parameter $\alpha /M^2 = 0.1$ is fixed. Solid curves represent stable solutions, while dotted lines correspond to unstable ones. The gray areas are non-physical.}
\label{Fig_sol}
\end{figure}

In our wormhole construction, we take metric functions $A_{1,2}$  given by Eq. (\ref{A-metric}), with the same values of mass $M$ and charge $Q$ but different values of the curvature $R_{1,2}$ at each side of the throat (which is equivalent to adopt different cosmological constants $\Lambda _{1,2}$). It is important to remind that we take a radius $a$ of the throat large enough to avoid the presence of the event horizon and the singularity at any side of the throat, and we also choose it smaller than the cosmological horizon when present. One expects that the term corresponding to the charge in Eq. (\ref{A-metric}) has the same sign as in General Relativity, so it is necessary that the inequality $F'(R)=1+2\alpha R>0$ is satisfied at both sides of the throat. In $F(R)$ gravity this inequality also implies a positive effective gravitational constant $G_{\mathrm{eff}}=G/F'(R)$ and the absence of ghosts \cite{bhfr2}. Then, in what follows we take values of $\alpha$ and $R_{1,2}$ so that $F'(R_{1,2})>0$. The possible radii $a_0$ of the static solutions are obtained from Eq. (\ref{CondEstatico}), with the corresponding metric functions $A_{1,2}$ both given by Eq. (\ref{A-metric}), with mass $M$, charge $Q$, and curvature values $R_{1,2}$. With the help of the definitions
\begin{equation}
\Omega_{1,2}=\left(\frac{2 M}{a_0^2} -\frac{2 Q^2}{(1+2 R_{1,2} \alpha )a_0^3} -\frac{R_{1,2}a_0}{6}\right) \left(1-\frac{2 M}{a_0}+\frac{Q^2}{(1+2 R_{1,2} \alpha ) a_0^2}-\frac{R_{1,2}a_0^2 }{12}\right)^{-1/2},
\label{Omega}
\end{equation}
\begin{equation}
\Phi _{1,2}=\frac{1}{a_0}\left( 1-\frac{2 M}{a_0}+\frac{Q^2}{(1+2 R_{1,2} \alpha ) a_0^2 }-\frac{R_{1,2}a_0^2 }{12}\right)^{1/2},
\label{Phi}
\end{equation}
for the charged wormholes Eq. (\ref{CondEstatico}) can be rewritten in the form
\begin{equation}
\Omega_1 + 4 \Phi _1 + \Omega_2 + 4 \Phi _2 = 0.
\label{CondEstaticoq}
\end{equation}
We show the results graphically in Fig. \ref{Fig_sol}, where we have selected the most representative ones. We have used in all plots $\alpha/M^2=0.1$, since different values do not alter significantly the general behavior of the solutions. The only effect that $\alpha/M^2$ has over the results is a small translation of the graphics along the charge axis. The values of the scalar curvature (in modulus) at each side of the throat used in the figures are $|R_{1}|M^2=0.2$ and $|R_{2}|M^2=0.4$, with the corresponding signs shown in each plot. The signs of $R_{1}$ and $R_{2}$ are more important than the values themselves, because they provide significant differences in the general behavior of the solutions. The gray regions have no physical meaning since they correspond to the inner zones of the event horizons of original manifolds, which were removed during the construction of the wormhole, or to the region inside the cosmological horizon (if present). The results show different behaviors around $Q_c/M$, where $Q_c=\mathrm{max}\{Q_c^1,Q_c^2\}$ is the critical charge, with $Q_c^{1,2}$ the values from which the inner and the event horizons of each original metric fuse into one. Depending on the values of the parameters, there are up to three solutions of Eq. (\ref{CondEstaticoq}); we number them from the smallest to the largest ones. The static solutions are stable when the sign of the second derivative of the potential is positive, i.e. when replacing the corresponding radius $a_0$ in Eq. (\ref{potencial2der}) gives $V''(a_0)>0$. In Fig. \ref{Fig_sol}, the stable and unstable solutions are displayed with solid and dotted lines, respectively. The behavior of these solutions depends on the different signs of $R_1$ and $R_2$. Considering this:
\begin{itemize}
\item In the cases when any of the curvature scalars is positive, i.e. $R_1>0$ or $R_2>0$, there are three solutions, two of them appear for a short range of $|Q|/M$, one is stable and the other is unstable. Both fuse into one for a particular value of $|Q|/M$, from which they are no longer present. The third solution  exists for any value of $|Q|/M$, it is unstable and close to the cosmological horizon.
\item When $R_1<0$ and $R_2<0$, there are two solutions for a small range of values of $|Q|/M$ with the same characteristics described above, one of them unstable and the other one stable, while the third unstable solution found in the previous case no longer exists. The stable solution is the one with the largest radius.
\end{itemize}
We can observe that in all cases shown in the plots, there is always a stable solution with radius $a_0/M$ for a restricted range of values of the quotient $|Q|/M$ between the charge and the mass. This range begins at $Q_c/M$ and ends shortly after.

\begin{figure}[t!]
\centering
\includegraphics[width=\textwidth]{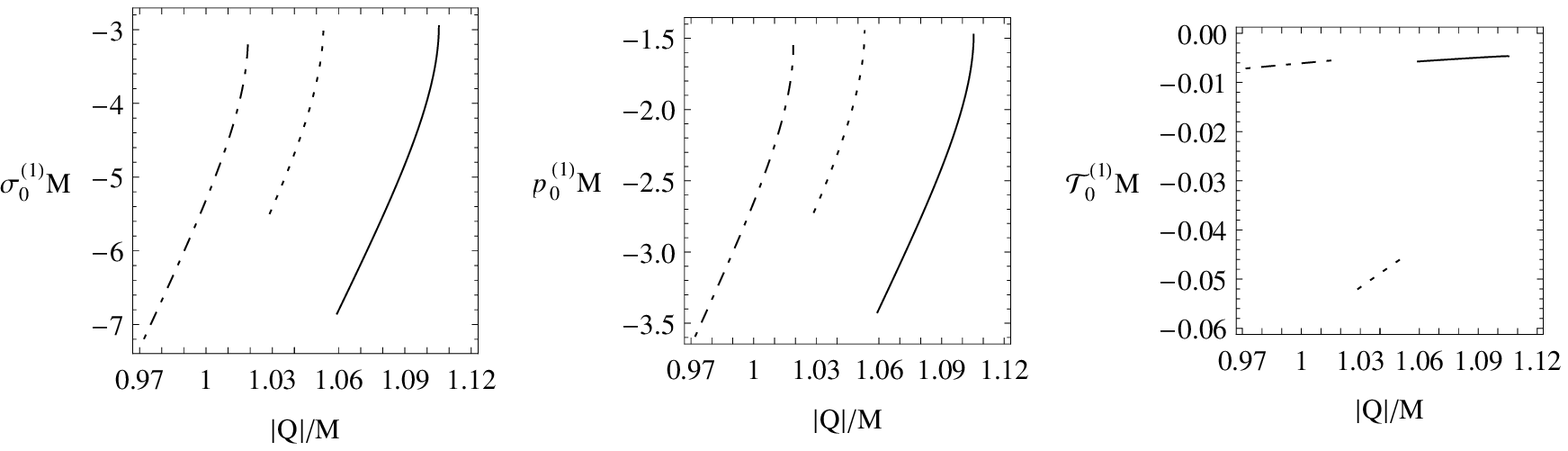}\\
\includegraphics[width=\textwidth]{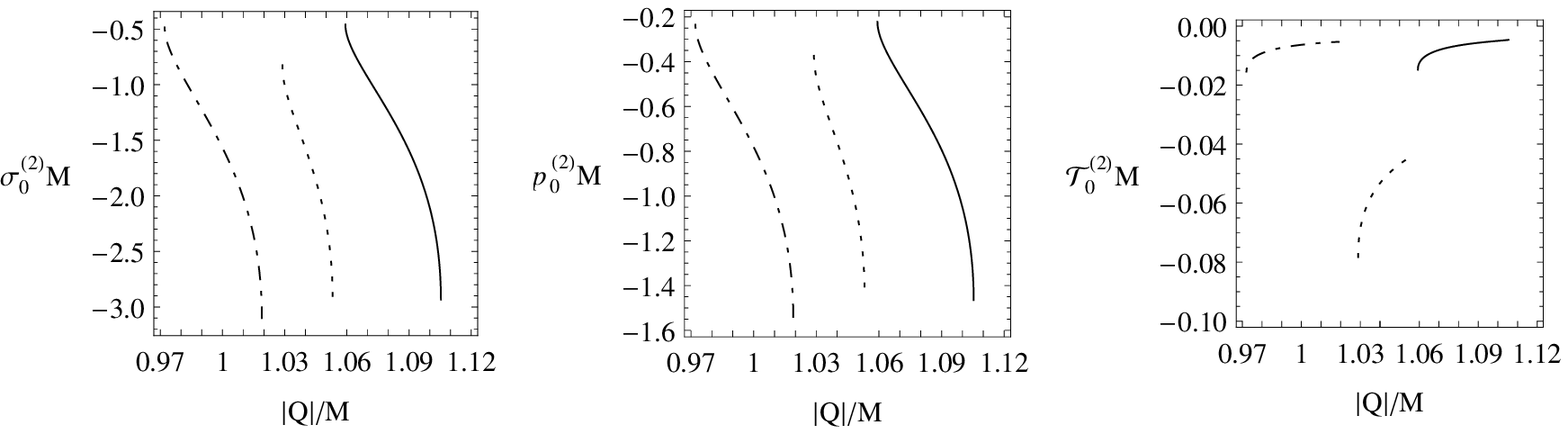}\\
\includegraphics[width=\textwidth]{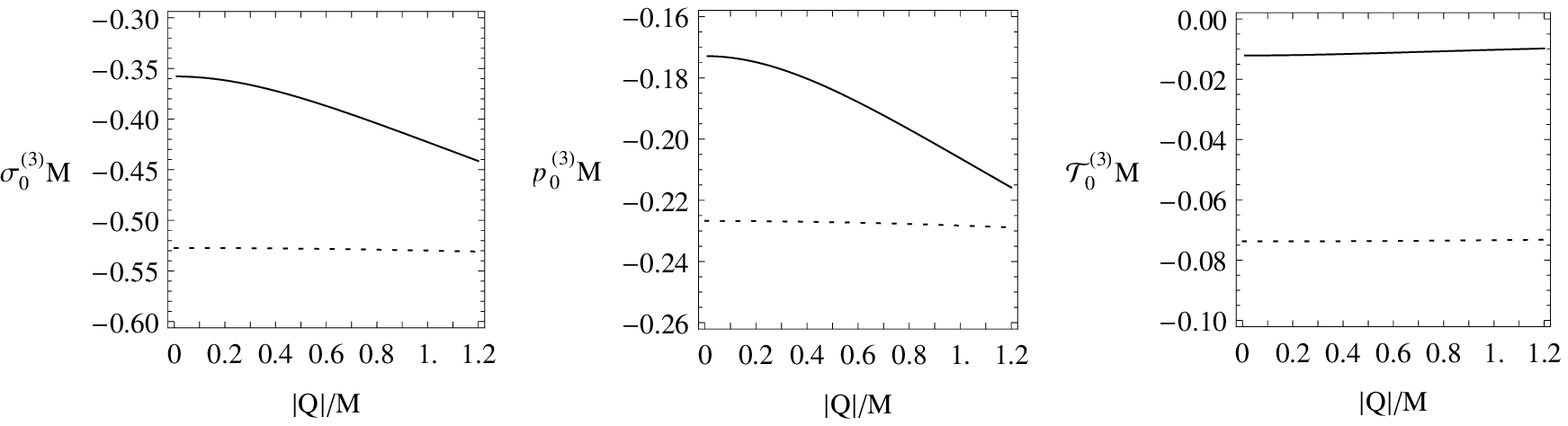}
\caption{The energy density $\sigma_0$, the pressure $p_0$, and the external tension $\mathcal{T}_0$ are plotted as functions of the quotient $|Q|/M$ between the absolute value of the charge and the mass. The parameter $\alpha /M^2 = 0.1$ is fixed. In all plots, where $|R_1| M^2=0.2$ and $|R_2| M^2=0.4$, the solid line corresponds to the $R_1>0$ and $R_2>0$ case, the dotted line to $R_1>0$ and $R_2<0$ case, and the dashed-dotted line to $R_1<0$ and $R_2<0$ case. Each row corresponds to a different static solution.}
\label{fig_ept}
\end{figure}

For the static solutions found above, by replacing the metric functions given by Eq.  (\ref{A-metric}) and using Eqs. (\ref{Omega}) and (\ref{Phi}), we obtain the energy density from Eq. (\ref{enden0}), the pressure from Eq. (\ref{pres0}), and the external tension from Eq. (\ref{T0}):
\begin{equation}
\sigma_0 =\frac{ \Omega_1 (1+2  \alpha R_1) + \Omega_2 (1+2 \alpha R_2 )}{2\kappa},
\label{enden0q}
\end{equation}
\begin{equation}
p_0 =-\frac{ \Phi _1(1+2 \alpha R_1) + \Phi _2 (1+2 \alpha R_2 )}{\kappa},
\label{pres0q}
\end{equation}
and
\begin{equation}
\mathcal{T}_0 = \frac{\alpha R_1 (\Omega_1 +4\Phi _1)+\alpha R_2 (\Omega_2+4 \Phi _2) }{\kappa}.
\label{T0q}
\end{equation}
The values of $\sigma_0$, $p_0$, and $\mathcal{T}_0 $ are shown as functions of $|Q|/M$ in Fig. \ref{fig_ept}. Again, we have taken $\alpha/M^2=0.1$, $|R_{1}|M^2=0.2$, and $|R_{2}|M^2=0.4$.  For the different combinations of signs of $R_1$ and $R_2$, we can summarize the results as follows:
\begin{itemize}
\item The absolute value of the (negative) energy density for the first solution decreases with $|Q|/M$, while the absolute values of the energy density for the second and third solutions increases with $|Q|/M$.
\item The absolute value of the (negative) pressure corresponding to the first solution decreases with $|Q|/M$, while for the second and third solutions increase with $|Q|/M$.
\item Regardless of the solution we are considering, the external tension is negative (so it is a pressure) and its absolute value decreases with $|Q|/M$ in any case. 
\end{itemize}
We see in all cases that $\sigma _0<0$ and $\sigma _0 + p_0<0$, so the weak and the null energy conditions are not satisfied, i.e. the fluid at the throat is exotic. 

The dual layer distribution strength, in the orthonormal basis, has components with absolute values $|\mathcal{P}_{\hat{\imath}\hat{\jmath} }|=2|\alpha [R]| \delta _{\hat{\imath}\hat{\jmath}} /\kappa$, which are independent of the mass and the charge. The dependence of the dual layer distribution $\mathcal{T}_{\hat{\imath}\hat{\jmath} }$ with $M$ and $Q$ is only through the unit normal and the covariant derivative.

\section{Conclusions}\label{conclu}

Within the framework of the quadratic $F(R)=R- 2\Lambda +\alpha R^2$ theory, we have constructed a class of spherically symmetric wormholes by cutting and pasting two manifolds with constant and different curvature scalars, into a hyper-surface representing the throat. In this way, the resulting thin--shell wormhole is not symmetric across the throat.  The construction forces a condition that determines the equation of state between the energy density $\sigma$, the pressure $p$, and the external tension $\mathcal{T}$ at the throat, which should satisfy  $\sigma-2p=\mathcal{T}$. There is also a double layer energy--momentum tensor distribution at the throat. The nature of this layer, resembling a dipole distribution of electromagnetism, seems to be unclear \cite{js1,js2,js3,js4}. Recently, it was shown that the presence of a double layer distribution is a shared feature of quadratic gravitational theories \cite{js4}. We have presented a general formalism for the analysis of the stability of static configurations under perturbations preserving the spherical symmetry. As a particular example, we have taken for the construction the outer parts of two manifolds with the same mass $M$ and charge $Q$, and with different constant values $R_1$ and $R_2$ of the curvature scalars. We have found the radius $a_0$ of the static solutions for different combinations of $R_1$ and $R_2$. When one of the curvature scalars is positive, or both, there exist three solutions: the two smaller ones, which appear for a short range of values of  $|Q|/M$, and the largest one present for any value of $|Q|/M$, which has a throat radius close to the cosmological horizon. In the case when both curvature scalars are negative, there exist only the first two solutions mentioned above; the third one is absent. The third solution and the smaller one of the other two are unstable under radial perturbations, while the remaining one is stable. We have found that the matter has a negative surface energy density at the throat, and therefore exotic matter is required for threading the wormhole. Regardless of the sign of the curvature scalars, we have obtained that the absolute value of the energy density of the first solution decreases in terms of $|Q|/M$, while the absolute value of the energy density of the second and third (when it exists) solutions increases. A similar behavior can be found related to the pressure. Considering the absolute value of the (negative) external tension, it decreases with $|Q|/M$ no matter which solution we are taking into account. The dual layer distribution strength has components in the orthonormal frame $\mathcal{P}_{\hat{\imath}\hat{\jmath} }$ which are proportional to the quadratic coefficient $\alpha$ and to the jump in the curvature $[R]$, and they do not depend on the mass and the charge, so the dual layer distribution tensor $\mathcal{T}_{\hat{\imath}\hat{\jmath} }$ dependence on these parameters is due to their presence in the unit normal and the covariant derivative.

\section*{Acknowledgments}

This work has been supported by CONICET and Universidad de Buenos Aires.

\end{document}